\documentclass[usenatbib]{mnras}
\usepackage[english]{babel}
\usepackage[utf8]{inputenc}
\usepackage[T1]{fontenc}
\usepackage{graphicx,epstopdf}
\usepackage{amsmath}
\usepackage{amssymb}
\usepackage{wasysym}
\title[The faint SN 2018hwm and its very long plateau]{Low luminosity Type II supernovae – III. SN 2018hwm, a faint event with an unusually long plateau}
\author[Reguitti et al.]{
Reguitti A.$^{1,2,3}$\thanks{E-mail: andreareguitti@gmail.com},
Pumo M. L.$^{4,1,5}$, Mazzali P. A.$^{6,7}$, Pastorello A.$^{1}$, Pignata G.$^{2,3}$, \newauthor
\enskip Elias-Rosa N.$^{1,8}$, Prentice S. J.$^{9}$, Reynolds T.$^{10}$, Benetti S.$^{1}$, Mattila S.$^{10}$,  \newauthor
\enskip Kuncarayakti H.$^{10,11}$, Rodrìguez O.$^{12}$\\
$^{1}$INAF – Osservatorio Astronomico di Padova, Vicolo dell'Osservatorio 5, 35122 Padova, Italy\\
$^{2}$Departamento de Ciencias F\'{i}sicas – Universidad Andrés Bello, Avda. Rep\'{u}blica 252, Santiago 8320000, Chile\\
$^{3}$Millennium Institute of Astrophysics, Nuncio Monsenor S\'{o}tero Sanz 100, Providencia, Santiago 8320000, Chile\\
$^{4}$Università degli studi di Catania, Dipartimento di Fisica e Astronomia ``E. Majorana'', Via S. Sofia 64, Catania, 95123, Italy\\
$^{5}$Laboratori Nazionali del Sud-INFN, Via S. Sofia 64, Catania, 95123, Italy\\
$^{6}$Astrophysics Research Institute, Liverpool John Moores University, ic2, 146 Brownlow Hill, Liverpool L3 5RF, UK\\
$^{7}$Max-Planck Institut fur Astrophysik, Karl-Schwarzschild-Str. 1, D-85741 Garching, Germany\\
$^{8}$Institute of Space Sciences (ICE, CSIC), Campus UAB, Carrer de Can Magrans s/n, 08193 Barcelona, Spain\\
$^{9}$School of Physics, Trinity College Dublin, The University of Dublin, Dublin 2, Ireland\\
$^{10}$Tuorla Observatory, Department of Physics and Astronomy, FI-20014 University of Turku, Finland\\
$^{11}$Finnish Centre for Astronomy with ESO (FINCA), FI-20014 University of Turku, Finland\\
$^{12}$School of Physics and Astronomy, Tel Aviv University, Tel Aviv 69978, Israel\\
}
\date{Accepted XXX. Received YYY; in original form ZZZ}
\begin{document}
\maketitle
\volume{00}
\pagerange{1-14}
\pubyear{2020}

\begin{abstract}
In this work, we present photometric and spectroscopic data of the low-luminosity Type IIP supernova (SN) 2018hwm. The object shows a faint ($M_r=-15$ mag) and very long ($\sim$130 days) plateau, followed by a 2.7 mag drop in the $r$-band to the radioactive tail. The first spectrum shows a blue continuum with narrow Balmer lines, while during the plateau the spectra show numerous metal lines, all with strong and narrow P-Cygni profiles. The expansion velocities are low, in the 1000-1400 km s$^{-1}$ range. The nebular spectrum, dominated by H$\alpha$ in emission, reveals weak emission from [O I] and [Ca II] doublets. The absolute light curve and spectra at different phases are similar to those of low-luminosity SNe IIP.
We estimate that 0.0085 $M_{\odot}$ of $^{56}$Ni mass were ejected, through hydrodynamical simulations. The best fit of the model to the observed data is found for an extremely low explosion energy of 0.075 foe, a progenitor radius of 845 $R_{\odot}$ and a final progenitor mass of 9-10 $M_{\odot}$.
Finally, we performed a modeling of the nebular spectrum, to establish the amount of oxygen and calcium ejected. We found a low M($^{16}$O)$\approx 0.02$ $M_{\odot}$, but a high M($^{40}$Ca) of 0.3 $M_{\odot}$.
The inferred low explosion energy, the low ejected $^{56}$Ni mass and the progenitor parameters, along with peculiar features observed in the nebular spectrum, are consistent with both an electron-capture SN explosion of a super-asymptotic giant branch star and with a low-energy, Ni-poor iron core-collapse SN from a 10-12 $M_{\odot}$ red supergiant.
\end{abstract}

\begin{keywords}
supernovae: general, supernovae: individual: SN 2018hwm
\end{keywords}

\section{Introduction} \label{introduction}
Type II supernovae (SNe) are luminous events that originate from the explosion of a massive ($>8 M_{\odot}$) star.
Historically, Type II SNe were subdivided in two main subclasses \citep{barbon} based on their photometric behaviour, namely Type IIL (with a linearly declining light curve after maximum) and Type IIP SNe. The latter represent events whose luminosity stays at a nearly constant luminosity for a long period, hence called `plateau' phase.
The long plateau of SNe IIP \citep[90-100 days,][]{hamuy} is due to the energy released by the hydrogen envelope as it cools and recombines, after being fully ionized by the explosion.
Much work has been done to characterize the properties of SNe IIP through the analysis of large samples of objects \citep{li2,arcavi,anderson,faran,gutierrez1,sanders,valenti,gutierrez2,gutierrez3}.
In particular, a small group of SNe IIP are remarkably less luminous than average. The first sub-luminous SN IIP discovered was SN 1997D \citep{turatto,benetti}, that had a peak luminosity of only $M_B=-14.65$ mag. During the nebular phase, the light curve of SN 1997D shows a tail with a declining slope compatible with the radioactive decay of $^{56}$Co, indicating that at this late time the luminosity of the object is solely powered by a small amount of radioactive material (2$\times 10^{-3}$ $M_{\odot}$).

SN 1999br \citep{zampieri2,pasto1} was very similar in luminosity and spectral appearance to SN 1997D, with narrow metal lines indicating low expansion velocities and unusually strong Ba II lines. In addition, SN 1999br showed a long plateau, lasting at least 110 days. A long plateau is indicative of a large recombination time, and may constitute evidence for the presence of a massive H envelope, that is consistent with the deduced massive ejecta \citep[14 $M_{\odot}$,][]{zampieri2}.

SN 2003Z \citep{knop,utrobin1,spiro2} was spectroscopically well followed among low-luminosity (LL) SNe IIP, with spectra that cover all evolutionary phases, from the early plateau, showing broad P-Cygni features, to the nebular phase.

The LL SN IIP SN 2005cs \citep{pasto2,pasto3} exploded in the nearby (8.4 Mpc) galaxy M 51. Because of the small distance, it was possible to directly observe the progenitor star in archive Hubble Space Telescope ($HST$) images, that appeared to be a Red Supergiant (RSG) with an initial mass in the 7–13 $M_{\odot}$ range \citep{maund1,li1,takats1}. The proximity of SN 2005cs also allowed observations of its faint radioactive tail, and it was obtained an ejected $^{56}$Ni mass of only 3$\times10^{-3}$ $M_{\odot}$ \citep{pasto3}.

\cite{galyam} conducted a multi-wavelength follow-up of the LL SN 2010id, from $\gamma$-rays to the radio. It was observed extremely early, with a spectrum taken only 1.5 days after explosion. The rise to the plateau phase lasted less than 2 days.

Another remarkable object was SN 2016bkv \citep{nakaoka}, which showed a very long plateau of 140 days, and an early bump in the light curve. A similar bump was likely observed also in SN 2003Z \citep{spiro2}. During the bump, the spectrum showed a blue continuum and a narrow H$\alpha$ emission, resembling those observed in Type IIn SNe, that indicate interaction between the SN ejecta and pre-existing circumstellar material (CSM). The presence of CSM suggests that the progenitor experienced mass loss just before the explosion.

The family of LL SNe IIP has been extended by the samples of \cite{pasto1} and \cite{spiro2}, and all objects share similar properties: low-luminosity, with absolute magnitudes $M_V$ during the plateau between $\sim-14$ and $\sim-15.5$ mag, compared to the average absolute magnitude of normal IIP SNe of $-16.7$ mag \citep{anderson}, narrow spectral lines indicating low expansion velocities of about 1000 km s$^{-1}$, and small ejected $^{56}$Ni masses ($<10^{-2}M_{\odot}$), one order of magnitude smaller than normal SNe \citep{benetti,spiro1}. Another remarkable property of LL SNe IIP is the presence of narrow-lined spectra during the plateau, although narrow spectral lines were seen also in transitional IIP objects.

The luminosity gap between normal and faint SNe IIP has been filled in fact by a few `transitional' objects including: SNe 2008in \citep{roy}, 2009N \citep{takats2}, 2009js \citep{gandhi}, 2013am \citep{zhang,tomasella2}, 2013K \citep{tomasella2} and 2018aoq \citep{oneill}. These objects create a continuous distribution of luminosity among Type IIP SNe. Whilst transitional SNe IIP are brighter than faint ones, the low expansion velocities inferred from the spectral lines resemble those observed in LL SNe IIP.

An extended survey conducted on pre-explosion images of normal Type IIP SNe have confirmed that Red Super-Giant (RSG) stars, with masses in the range 8-16 $M_{\odot}$, are the most plausible progenitors of those objects (\citealt{smartt1} and references therein, \citealt{vandyk3}, \citealt{maund2}, \citealt{smartt3}).
A low mass, RSG star of 8-8.5 $M_{\odot}$, with a radius of 500 $R_{\odot}$, was identified as the progenitor of SN 2008bk (\citealt{mattila1}, \citealt{vandyk2}, \citealt{maund3}, Pignata et al., in preparation), and of SN 2003gd (8-9 $M_{\odot}$, \citealt{vandyk1}, \citealt{smartt0}). Evolutionary numerical simulations of a Main Sequence (MS) 12 $M_{\odot}$ star reproduce the observed features of SN 2008bk, confirming a low-mass supergiant as the progenitor \citep{lisakov1}.
\cite{fraser} calculate the most probable progenitor mass range for the subluminous Type IIP SNe to be between 7.5 and 9.5 $M_{\odot}$, hence LL SNe IIP could arise from the lower-mass-end of RSG or Super-Asymptotic Giant Branch (SAGB) stars.

The properties of the progenitors of SNe, such as mass and radius, can be derived also by modelling light curves and spectra, and through hydrodynamical simulations \citep{chugai,utrobin1,utrobin2,bersten,morozova}.
The simulations indicate that some LL SNe IIP come from low-to-moderate mass and large radii progenitors (250-500 $R_{\odot}$, 10-11 $M_{\odot}$ ejecta, 13-15 $M_{\odot}$ on MS), with low explosion energies \citep[$\sim$10$^{50}$ erg,][]{pumo4}.

More recently, another promising way to derive physical characteristics of the progenitor has been proposed by modeling nebular spectra (e.g. \citealt{jerkstrand1,jerkstrand2,jerkstrand3}). The presence/absence and the strength of some specific lines, such as the [O I] and [Ca II] doublets, give important clues on the initial mass and progenitor's type.

In this context, in this paper we present a study of SN 2018hwm, a new LL SN IIP exploded in the relatively nearby galaxy IC 2327, and with a very long plateau.
The paper is structured as follows: In Sect. \ref{discovery} the discovery of SN 2018hwm and the properties of its host galaxy are reported. In Sect. \ref{photometry} we present and analyse the photometric data. The spectra are presented in Sect. \ref{spectroscopy}. In Sect. \ref{discussion} we discuss the nature of the progenitor and the explosion scenario of SN2018hwm on the basis of the SN's observables.

\section{Discovery and host galaxy} \label{discovery}
\begin{figure}
\includegraphics[width=\columnwidth]{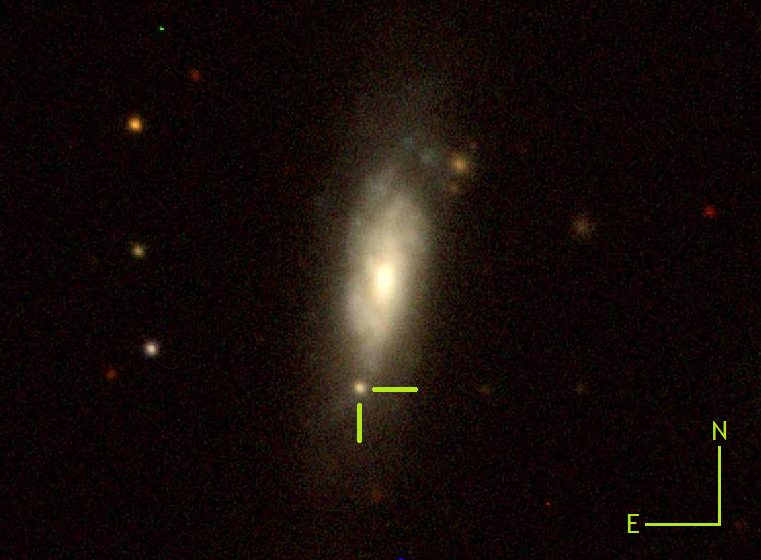}
\caption{Colour image of the host galaxy IC 2327 and of SN 2018hwm taken on 2019 February 26, 4 months after explosion, with the Liverpool Telescope \citep{steele}. The image is a combination of the frames obtained with the $g$, $r$ and $i$ filters. The location of the SN is marked.}
\label{fig1}
\end{figure}

SN 2018hwm (a.k.a. ZTF18acurqaw, ATLAS18zrw, PS18byn) was discovered by the Puckett Observatory Supernova Search \citep[POSS\footnote{The POSS survey utilizes a 24-inches telescope, with an Apogee U47 CCD, located at Mayhill, NM, USA.};][]{gagliano}, on 2018 November 4.51 (UT) in the galaxy IC 2327 (or UGC 4356), at celestial coordinates $\alpha$ = 08:21:28.192, $\delta$ = +03:09:52.35 (J2000). At discovery, the unfiltered magnitude of the transient was 19.2.
A colour image of the host galaxy with the SN taken 4 months after the discovery, and obtained combining the frames in the $g$, $r$ and $i$ filters, is shown in Figure \ref{fig1}.

The spectroscopic classification of SN 2018hwm was performed 4 days after discovery by the extended Public ESO Spectroscopic Survey for Transient Objects \citep[ePESSTO;][]{smartt2} collaboration with the ESO 3.58-m New Technology Telescope (NTT). The transient was classified as a young type II SN, showing a blue continuum and P-Cygni Balmer lines \citep{congiu}.
The tools used for the classification were \texttt{GELATO} \citep{harutunyan} and \texttt{SNID} \citep{blondin}. The classification spectrum is presented in Sect. \ref{spectroscopy}. Later, the transient developed the typical features of SNe IIP (see Sect. \ref{spectroscopy}).

IC 2327 is an Sa spiral galaxy according to \cite{devaucouleurs}. 
The NASA/IPAC Extragalactic Database (NED\footnote{https://ned.ipac.caltech.edu}) reports different measurements of distance of IC 2327, all obtained through the Tully-Fisher relation method \citep[e.g.][]{tully}. We adopt the weighted mean value of $52\pm5$ Mpc, assuming a standard cosmology ($H_0=73$ km s$^{-1}$Mpc$^{-1}$, $\Omega_m=0.27$, $\Omega_{\Lambda}=0.73$). This translates in a distance modulus (DM) of 33.58$\pm$0.19 mag.
The redshift of the host galaxy $z=0.00895\pm0.00002$ is from \cite{falco}.
The Milky Way reddening in the direction of IC 2327 is $A_{V,MW}=0.071$ mag \citep{schlafly}.
The extinction contribution of the host galaxy is negligible, as in the spectra (Sect. \ref{spectrallines}) we do not see absorption lines from the Na I D doublet at the host galaxy redshift, that would suggest the presence of additional dust.

In order to understand the nature of the progenitor of SN 2018hwm, we evaluated some properties of the host galaxy, including the metallicity and the star formation rate. We estimated the metallicity of the host galaxy through the correlation between the mean [O/H] metallicity indicator and the $B$-band absolute magnitude of the galaxy \citep{pilugin}. From the HyperLeda\footnote{http://leda.univ-lyon1.fr/} database, IC 2327 has $M_B=-19.93\pm0.42$ mag. Using equation 12 from Pilyugin et al. 2004 for spiral galaxies, we obtain 12+log[O/H]$_{host}$=8.50$\pm0.77$. Assuming a solar value of 12+log[O/H]$_{\odot}$=8.69 \citep{asplund}, this corresponds to $Z=0.013$ (adopting $Z_{\odot}=0.02$), thus we may conclude that the oxygen abundance of the host is nearly solar.

To estimate the star formation rate (SFR) in IC 2327, we use the \cite{kennicutt} relation between the SFR and the luminosity in the far ultraviolet (FUV) region: $SFR(FUV,M_{\odot}yr^{-1})=1.4\times10^{-28}L(FUV,erg\cdot s^{-1} Hz^{-1})$. The NED database reports two integrated flux densities in the FUV (at 1530 \AA) from \textit{GALEX}, of 1.11 and 1.15 mJy. Taking the average flux density of 1.13 mJy and the assumed distance of 52 Mpc, we obtain a SFR of 0.51 $M_{\odot}yr^{-1}$ for IC 2327.

\section{Photometry} \label{photometry}
The photometric follow-up of SN 2018hwm was performed with a plethora of instruments and telescopes, available to our collaborations, whose characteristics are reported in Table \ref{tab1}. Optical observations were done with Johnson-Cousins $BVRI$ and Sloan $griz$ filters, and in the near infrared (NIR) with $JHKs$ filters, albeit only for a few epochs.

\begin{table}
\caption{Observational facilities and instrumentation used in the photometric follow-up of SN 2018hwm.}
\label{tab1}
\begin{tabular}{llll}
\hline
Telescope & Location & Instrument & Filters \\
\hline
NOT (2.56m) & La Palma & ALFOSC & $BVgriz$ \\
NOT (2.56m) & La Palma & NOTCAM & $JHKs$ \\
LT (2.0m) & La Palma & IO:O & $BVgriz$ \\
NTT (3.58m) & La Silla & EFOSC & $V$ \\
TRAPPIST (0.6m) & La Silla & Fairchild & $BVRI$ \\
Oschin (1.22m) & Mt. Palomar & ZTF & $gr$ \\
SMARTS (1.3m) & CTIO & ANDICAM & $BVRI$ \\
PROMPT3 (0.6m) & CTIO & Apogee & $iz$ \\
PROMPT5 (0.4m) & CTIO & Apogee & $griz$ \\
Blanco (4.0m) & CTIO & DECAM & $gr$ \\
GTC (10.4m) & La Palma & OSIRIS & $r$ \\
\hline
\end{tabular}
\end{table}

For the photometric data reduction we used a dedicated pipeline called \texttt{SNOoPy}\footnote{SNOoPy is a package for SN photometry using PSF fitting and/or template subtraction developed by E. Cappellaro at the Padova Astronomical Observatory. A package description can be found at http://sngroup.oapd.inaf.it/snoopy.html.} \citep{cappellaro}. An exhaustive description of the reduction procedures can be found in \cite{reguitti}. The instrumental magnitudes are determined through the PSF-fit method. For Sloan filter images, the photometric zero points and colour terms were computed through a sequence of reference stars from the \textit{SDSS} survey in the SN field. For Johnson $BV$ filter frames, magnitudes for those reference stars were taken from the APASS DR10 catalogue\footnote{https://www.aavso.org/apass}.
Finally, for NIR images, the magnitudes were calibrated with the 2MASS catalogue \citep{skrutskie}.
Photometric errors were estimated through artificial star experiments, also accounting for uncertainties in the PSF-fitting procedure and the colour terms.

The observed optical Sloan, Johnson and NIR magnitudes are listed in Table \ref{tab4}, \ref{tab5} and \ref{tab6}, respectively.

\subsection{Light curve evolution} \label{lightcurves}
\begin{figure}
\includegraphics[width=1.1\columnwidth]{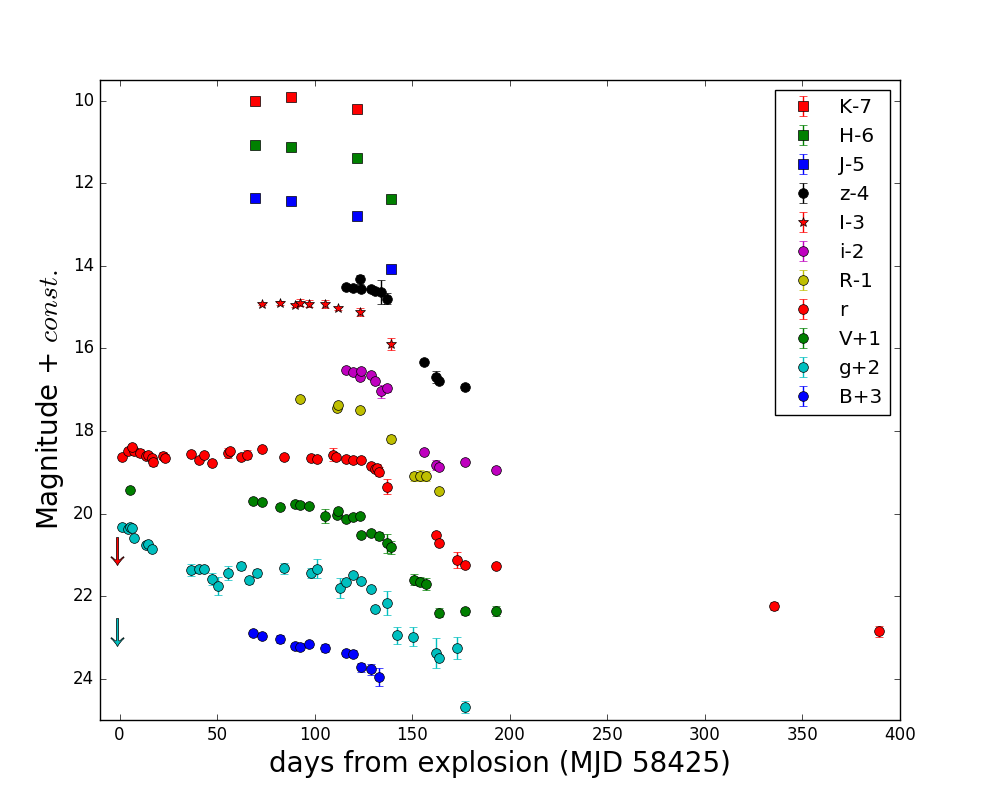}
\caption{Multiband ($BVRIgrizJHK$) light curves of SN 2018hwm, covering 400 days of evolution. The phases are relative to the assumed explosion epoch (MJD 58425.0$\pm$1.5). For clarity the curves of different filters are shifted by a constant.}
\label{fig2}
\end{figure}

The light curves of SN 2018hwm are plotted in Figure \ref{fig2}.
The $r$-band light curve is the best sampled, and it is used as a reference for studying the photometric evolution of SN 2018hwm. The Zwicky Transient Facility \citep[ZTF; \citealt{bellm},][]{graham} survey reports the last non-detection (at 20.9 mag) on 2018 November 1 (MJD 58423.5), while the first detection of the transient is reported 3 days later, on MJD 58426.5, nearly contemporary to the POSS discovery \citep{gagliano}. Because of this, we adopt MJD 58425.0$\pm$1.5 as the most likely explosion epoch. 
The maximum light is reached about 5.5 days after explosion. The maximum is followed by the plateau phase, which ends at +130 d from explosion, comparable to that of SN 2009ib \citep{takats3}, which is longer than for normal SNe IIP, whose plateau generally lasts 90-100 days \citep{hamuy}. During the plateau, the magnitude of the object remains nearly constant at $r=18.5\pm0.1$ mag. The post-plateau decline lasts around 1 month, and finally the SN sets into the radioactive slope at $r=21.2\pm0.2$ mag, resulting in a drop of $\sim$2.7 mag, which is quite common in SNe IIP, see \cite{olivares}. A similar drop of about 2.5 mag is observed also in the $V$ light curve. Then, the SN went into solar conjunction, and when visible again, was recovered at 22.2$\pm$0.1 mag (phase +335 d).

In the $g$-band, after the maximum, the light curve declines with a rate of $2.6\pm0.2$ mag (100 d)$^{-1}$ for about 40 days before reaching the plateau, that lasts until +120 d.
In the other bands, the plateau and the following drop are only partially covered. 
The $B$-band follow-up started only 2 months after maximum; between +60 and +115 d the light curve in the $B$ filter does not show a plateau, but a linear decline with a slope of $0.96\pm0.02$ mag (100 d)$^{-1}$.

Two additional photometric data points were collected in the $r$-band, about one year after explosion. Between +170 d and +390 d the observed decline slope is $0.74\pm0.06$ mag (100 d)$^{-1}$, slower than the rate expected from the $^{56}$Co radioactive decay (0.98 mag (100 d)$^{-1}$). Such a flattening during the early nebular phase has been previously observed by \cite{pasto1} in SN 1999eu.
However, we note that between the 2 final detections, the decline rate is $1.1\pm0.3$ mag (100 d)$^{-1}$, consistent with the value of the $^{56}$Co decay.

The duration of the plateau of SN 2018hwn ($\sim$130 d) is unusually long with respect that that observed in normal Type IIP SNe. \cite{anderson} conducted a study of 116 SNe IIP, and found that the distribution of the optically thick phase duration (OPTd), i.e. the time between the explosion and the end of the plateau, has a mean value of 84 days, with a dispersion of 17 days. The largest OPTd in their sample is observed for SN 2004er, with a duration of 120 d. Larger OPTds are also derived for the sub-luminous events (see also \citealt{valenti}). \cite{arcavi} and \cite{faran} found that the peak of the plateau duration distribution is around 90-100 days. If one considers the time between the explosion and the mid point of the transition from the ‘plateau’ to the radioactive tail ($t_{PT}$ in \citealt{anderson}), for SN 2018hwm this is 150 days, while \cite{sanders} for normal SNe IIP found values in the range 60-140 days, with a median at around 110 days. Again, the longest $t_{PT}$ is observed for SN 2004er, and is similar to that of SN 2018hwm.

\subsection{Absolute light curve}
We constructed the absolute $r$-band light curve of SN 2018hwm, adopting the Galactic reddening and DM from Sect. \ref{discovery}.
With these assumptions, the average absolute magnitude of SN 2018hwm during the plateau is $M_r=-15.0\pm0.2$ mag.
During the plateau the object was fainter in $g$-band, with the absolute magnitude staying constant at -14 mag between +40 and +120 days.
After the end of the plateau, the $r$ absolute magnitude of SN 2018hwm dropped to -12.5 mag. A similar value is reached also in the other bands (-12.2 in $g$, -12.3 in $V$), whereas in redder filters it remains a bit more luminous (-13.0 in $i$, -13.2 in $z$). 
During the radioactive tail the object further weakened, with last detection (+389 d) being at an absolute magnitude $M_r=-10.8\pm0.2$.

\begin{figure}
\includegraphics[width=1.1\columnwidth]{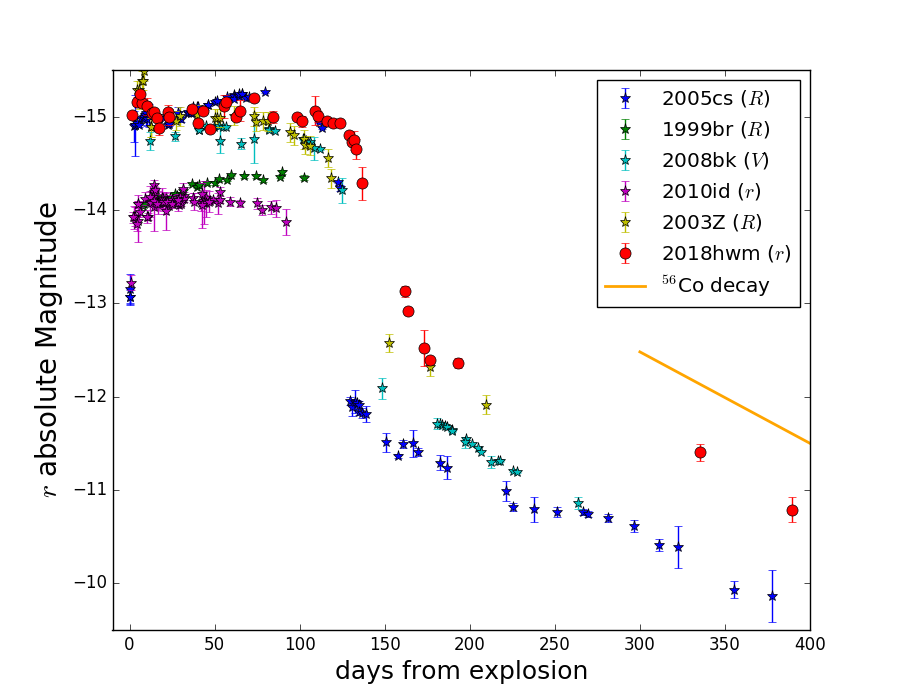}
\caption{$r$-band absolute light curve of SN 2018hwm compared to absolute light curves of LL IIP objects: SNe 1999br (Pastorello et al. 2004), 2005cs (Pastorello et al. 2009), 2008bk (Van Dyk et al. 2012), 2003Z (Spiro et al. 2014) and 2010id (Gal-Yam et al. 2011). Distance moduli and reddenings of the comparison objects are taken from their respective papers. The expected decay slope of $^{56}$Co (0.98 mag (100 d)$^{-1}$) is plotted for comparison.}
\label{fig3}
\end{figure}

We compared the absolute $r$-band light curve of SN 2018hwm with those of other known faint Type IIP SNe, i.e. SNe 1999br \citep{pasto1}, 2005cs \citep{pasto3}, 2008bk \citep{vandyk2}, 2003Z \citep{spiro2} and the particularly faint 2010id \citep{galyam}.
The distance modulus and reddening of each object are taken from the respective papers, and are rescaled to $H_0=73$ km s$^{-1}$Mpc$^{-1}$.
As can be seen in Figure \ref{fig3}, the light curve of SN 2018hwm is similar in  luminosity to those of SNe 2005cs, 2003Z and 2008bk, both at peak and during the plateau. However, the plateau phase in SN 2018hwm lasts nearly one month more than in other faint Type IIP SNe. The drop from the plateau begins around +130 days after explosion. In addition, the drop from the plateau in SN 2018hwm is smaller: SN 2018hwm is $\sim 1$ mag brighter than SNe 2005cs and 2008bk during the nebular phase (between 6 and 12 months after explosion).

In the context of large samples of SNe IIP, LL IIP are placed on the faint tail of a continuous distribution in luminosity. The sample of \cite{anderson} is characterized by a mean absolute magnitude at peak of $M_{V,max}=-16.7$ mag, with a scatter of 1 mag. Differently, \cite{li2} obtained an even fainter average absolute magnitude of $-16.1$ mag for a sample of Type II SNe, because of a different selection criteria of the sample.
While SN 2018hwm, with a maximum $M_r$ of -15.2 mag (see Figure \ref{fig3}), can be considered a faint object with respect to the global population of SNe IIP, it is a relatively luminous event in the context of the faint SNe IIP subgroup \citep{pasto1,spiro2}.

\section{Spectroscopy} \label{spectroscopy}
The spectroscopic monitoring of SN 2018hwm lasted about 1 year, during which we collected 6 optical spectra (see Table \ref{tab2}). All the spectra will be publicly released on the \textsc{WISeREP} repository \citep{yaron}.

\begin{table*}
\caption{Log of the spectroscopic observations of SN 2018hwm. For each spectrum, the date, the observed spectral range, the resolution, the exposure time and the telescope+instrument used are listed. The reported phases are relative to the assumed explosion time (MJD 58425.0, see Sect. \ref{lightcurves}).}
\label{tab2}
\begin{tabular}{lllllll}
\hline
Date & MJD & Phase & Coverage & Resolut. & Exposure & Telescope + \\
 & & (d) & (\AA) & (\AA) & (s) & Instrument \\
\hline
2018/11/08 & 58430.33 & +5.3   & 3640-9230 & 18 & 600  & NTT+EFOSC2 \\
2018/12/17 & 58468.44 & +43.4  & 3950-9220 & 25 & 2250 & P200+SEDM \\
2019/01/08 & 58491.04 & +66.0  & 3620-9640 & 19 & 2400 & NOT+ALFOSC \\
2019/01/15 & 58498.02 & +73.0  & 3740-9640 & 18 & 2500 & NOT+ALFOSC \\
2019/02/21 & 58535.97 & +111.0 & 3850-9640 & 18 & 2700 & NOT+ALFOSC \\
2019/10/27 & 58814.20 & +389.2 & 5100-10400 & 6 & 1380 & GTC+OSIRIS \\
\hline
\end{tabular}
\end{table*}

The NTT classification spectrum, taken through the ePESSTO program, was reduced with a \texttt{pyraf}-based pipeline \citep[\texttt{PESSTO},][]{smartt2}, optimised for the \textsc{EFOSC2} instrument.
The reduction operations performed by the pipeline include standard procedures: correction for bias and flat-field, extraction of the 1-D spectrum, removal of sky lines and cosmic rays, wavelength and flux calibrations, using arc lamps and spectrophotometric standard stars. 

The ZTF survey collected one low-resolution spectrum with the 200-inch Mount Palomar telescope and the SED Machine spectrograph \citep[SEDM;][]{blagonova}, that was reduced using the \texttt{pysedm} \citep{rigault} automatic pipeline.
The spectra from the NOT telescope were reduced using the \texttt{ALFOSCGUI}\footnote{https://sngroup.oapd.inaf.it/foscgui.html} pipeline \citep{cappellaro}, designed specifically for quickly reducing photometric and spectroscopic images taken with the \textsc{ALFOSC} instrument, as part of the NOT Unbiased Transients Survey (NUTS) collaboration \citep{mattila2}.
The nebular spectrum from GTC was reduced with routine IRAF procedures.

\begin{figure*}
\includegraphics[width=2\columnwidth]{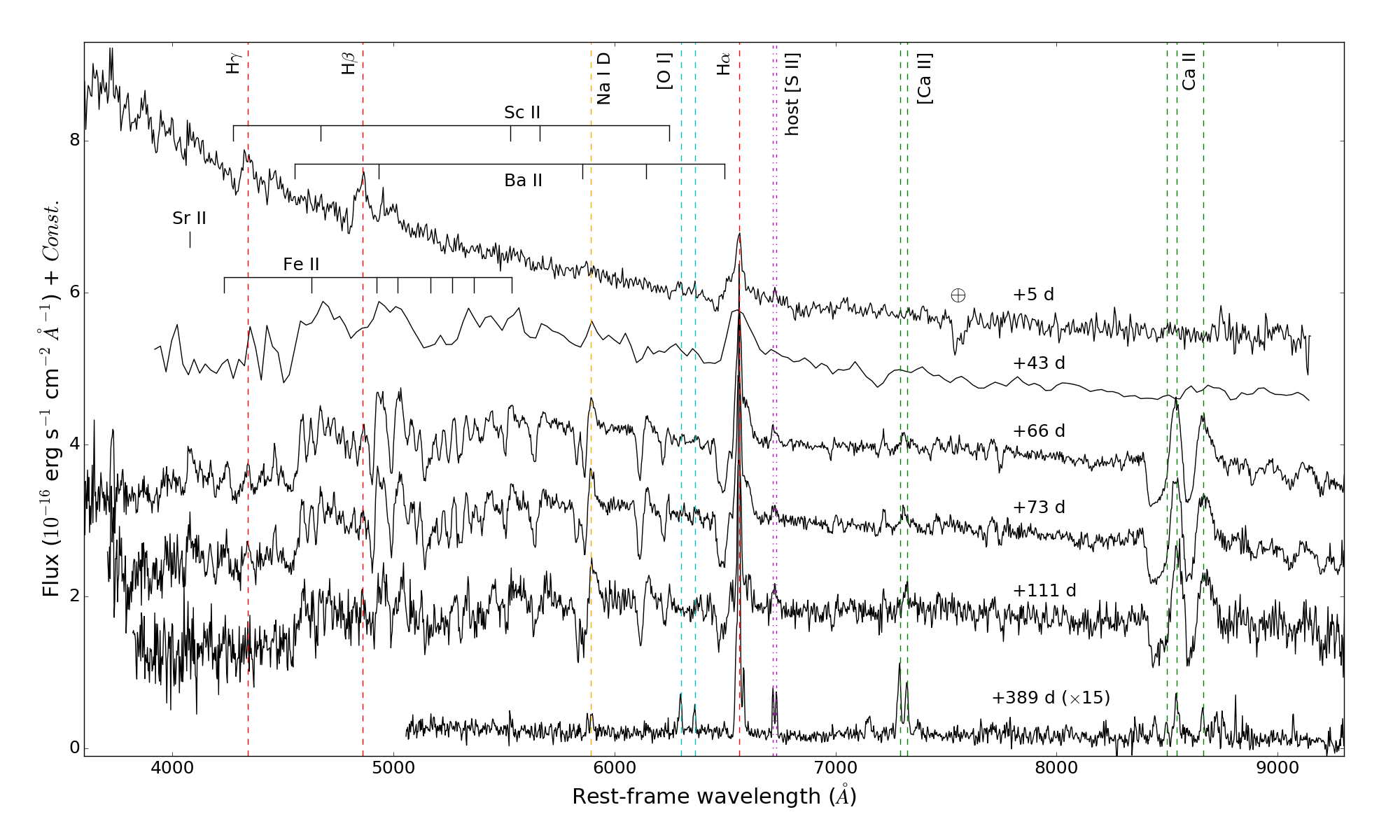}
\caption{The six spectra of SN 2018hwm, redshift- and reddening-corrected. The spectra are shifted in flux by a constant for clarity. The principal identified lines are marked. Heavy elements transitions are marked by short vertical lines, grouped by element. The strong telluric absorption band, not fully corrected in the ePESSTO spectrum, is also marked. The [S II] lines are marked differently, as they are interpreted as host galaxy contamination.}
\label{fig4}
\end{figure*}

The final spectra are corrected for the strongest telluric absorption bands, for redshift and for Galactic reddening using the \cite{cardelli} extinction law, and calibrated to match the closest broad-band photometry.
The six spectra and the identified lines are shown in Figure \ref{fig4}.

\subsection{Spectral evolution and line identification} \label{spectrallines}
The first spectrum, taken at around the maximum light, shows a blue continuum (with a black body temperature of 12000 K), and Balmer lines that have a relatively broad P-Cygni component, with a minimum blue-shifted by 4500 km/s, and a more prominent emission component, with $v_{FWHM}\sim$4000 km/s. Prominent, unresolved narrow emission lines from [O III] and [S II], superimposed to the broad components, are due to host galaxy contaminating sources \citep{congiu}.

We used the tools \texttt{SNID} \citep{blondin} and \texttt{GELATO} \citep{harutunyan} to spectroscopically establish the explosion epoch, based on a comparison between the first spectrum of SN 2018hwm and early spectra of other Type IIP SNe, with well-determined explosion time estimation, and searching for the most similar spectra.
The software found a good match with the spectra of SN 2005cs taken between 4 and 8 days after the explosion, consistent with our assumption about the explosion epoch of SN 2018hwm.

Although the second spectrum is of a lower quality than the first, absorption features from once ionized metal elements \citep[including Fe II, Ba II and Sc II, see the identification of][]{pasto1} start to emerge from a colder continuum ($T_{bb}\sim $ 6900 K).

We collected three spectra during the plateau phase, that show the typical features of Type IIP SNe. The three spectra, taken between 2 and 4 months after explosion, do not show a significant evolution, apart from a slow and modest decrease of the continuum temperature, with $T_{bb}$ decreasing from 4800 to 4500 K.
Together with H$\alpha$ and H$\beta$, we identify numerous lines coming from transitions of metals.
All the lines present an evident and narrow P-Cygni absorption. We measure the expansion velocity of the ejecta from the position of the minimum of the P-Cygni component in the three NOT spectra, and the values are reported in Table \ref{tab3}.

In the first photospheric spectrum the average veocity is 1400 km s$^{-1}$, while it decreases to 1300 km s$^{-1}$ in the second, and to 1000 km s$^{-1}$ in the third spectrum. The low velocities are an indication of a low explosion energy.
The metal absorption lines in the spectra of SN 2018hwm are very narrow, resembling those observed in transitional objects between normal and faint Type IIP SNe, like SNe 2013K and 2013am \citep{tomasella1}.
In the NIR part of the late spectra, two strong lines from the Ca II NIR triplet are detected.

The expansion velocity of the ejecta are relatively low in SN 2018hwm at all phases in comparison with the typical values observed in SNe IIP. \cite{gutierrez2} studied a sample of 122 SNe IIP, to infer the observed parameters of this class of objects. The ejecta expansion velocities at phase +50 d, measured from the FWHM of the H$\alpha$ line, are distributed between 9600 and 1500 km s$^{-1}$, with a median value of 7300 km s$^{-1}$. Around +100 d the velocities have decreased, but are still found within the 3000 to 7000 km s$^{-1}$ range. There are some noticeable outliers in the general distribution: SNe 2009aj and 2009au show low velocities, but are brighter than faint SNe IIP, and are known as luminous low expansion velocities SNe (LLEV; \citealt{rodriguez}). For SN 2018hwm we obtained a mean ejecta velocity of 1400 km s$^{-1}$ in the +66 d spectrum, and about 1000 km s$^{-1}$ at +111 d.

\begin{table}
\caption{Expansion velocities of the ejecta, measured from the position of the minimum of the P-Cygni profile of hydrogen and heavy metal lines, as observed in the three NOT spectra during the plateau phase. All velocities are in km s$^{-1}$. The typical error is of the order of $\pm50$ km s$^{-1}$.}
\label{tab3}
\begin{tabular}{lllll}
\hline
line & wavelength & spectrum 1 & spectrum 2 & spectrum 3 \\
 & (\AA) & +66 d & +73 d & +111 d \\
\hline
Sr II & 4078 & - & 1620 & - \\
Fe II & 4233 & - & 1630 & 1270 \\
Sc II & 4273 & - & 1050 & - \\
Ba II & 4554 & - & 1120 & 990 \\
Fe II & 4629 & 1420 & 1170 & - \\
Sc II & 4670 & 1540 & 1350 & 900 \\
H$\beta$ & 4861 & 1480 & 1170 & 1110 \\
Fe II & 4924 & 1280 & 1280 & 730 \\
Ba II & 4934 & - & - & 790 \\
Fe II & 5018 & 1610 & 1610 & 900 \\
Fe II & 5169 & 1450 & 1570 & - \\
Fe II & 5267 & 1020 & 970 & 970 \\
Fe II & 5363 & - & - & 890 \\
Sc II & 5527 & 1360 & 1520 & 1030 \\
Fe II & 5535 & 1490 & 1460 & 970 \\
Sc II & 5663 & 1380 & 1270 & - \\
Ba II & 5854 & 1330 & 1380 & 970 \\
Ba II & 6142 & 1320 & 1370 & 1180 \\
Sc II & 6246 & 1340 & 1100 & 910 \\
Ba II & 6497 & 1250 & - & 1200 \\
H$\alpha$ & 6563 & 1550 & 1370 & 1050 \\
\hline
\end{tabular}
\end{table}

Finally, we secured a nebular spectrum of SN 2018hwm about 1 year after the explosion. At that time, the transient had an apparent magnitude $r\approx$22.8 mag, but thanks to the high sensitivity of the GTC+OSIRIS combination, the spectrum has good signal-to-noise ratio, and shows some interesting features.
The spectrum is contaminated by the host galaxy contribution in the form of narrow (FWHM$\sim$7 \AA) emission lines, that are narrower than the SN features (with a mean FWHM of 13 \AA).
Over a flat pseudo-continuum we identify SN emission lines of [O I] $\lambda\lambda$6300,6364, H$\alpha$, [Fe II] $\lambda$7155, [Ca II] $\lambda\lambda$7291,7323 and the 3 lines of the NIR Ca II triplet $\lambda\lambda\lambda$ 8498,8542,8662. A faint blend of He I $\lambda$5876 and Na I D $\lambda\lambda$5890,5896 are also present. We also identify contaminant emission lines from the host, like [N II] $\lambda$6584, H$\alpha$ and [S II] $\lambda\lambda$6716,6731.
In particular, the H$\alpha$ line is made of 2 components: one from the SN, with a $v_{FWHM}$ of 600 km s$^{-1}$, and one stronger and narrower ($v_{FWHM}$=400 km s$^{-1}$, comparable to the spectral resolution of 300 km s$^{-1}$), from the host galaxy. The H$\alpha$ component from the SN is blue-shifted by 550 km s$^{-1}$, but a contribution from the host [N II] $\lambda$6548 line cannot be ruled out.
The O I $\lambda$7774 line is not detected, whereas the O I $\lambda$8446 line is visible.

\begin{figure*}
\includegraphics[width=2\columnwidth]{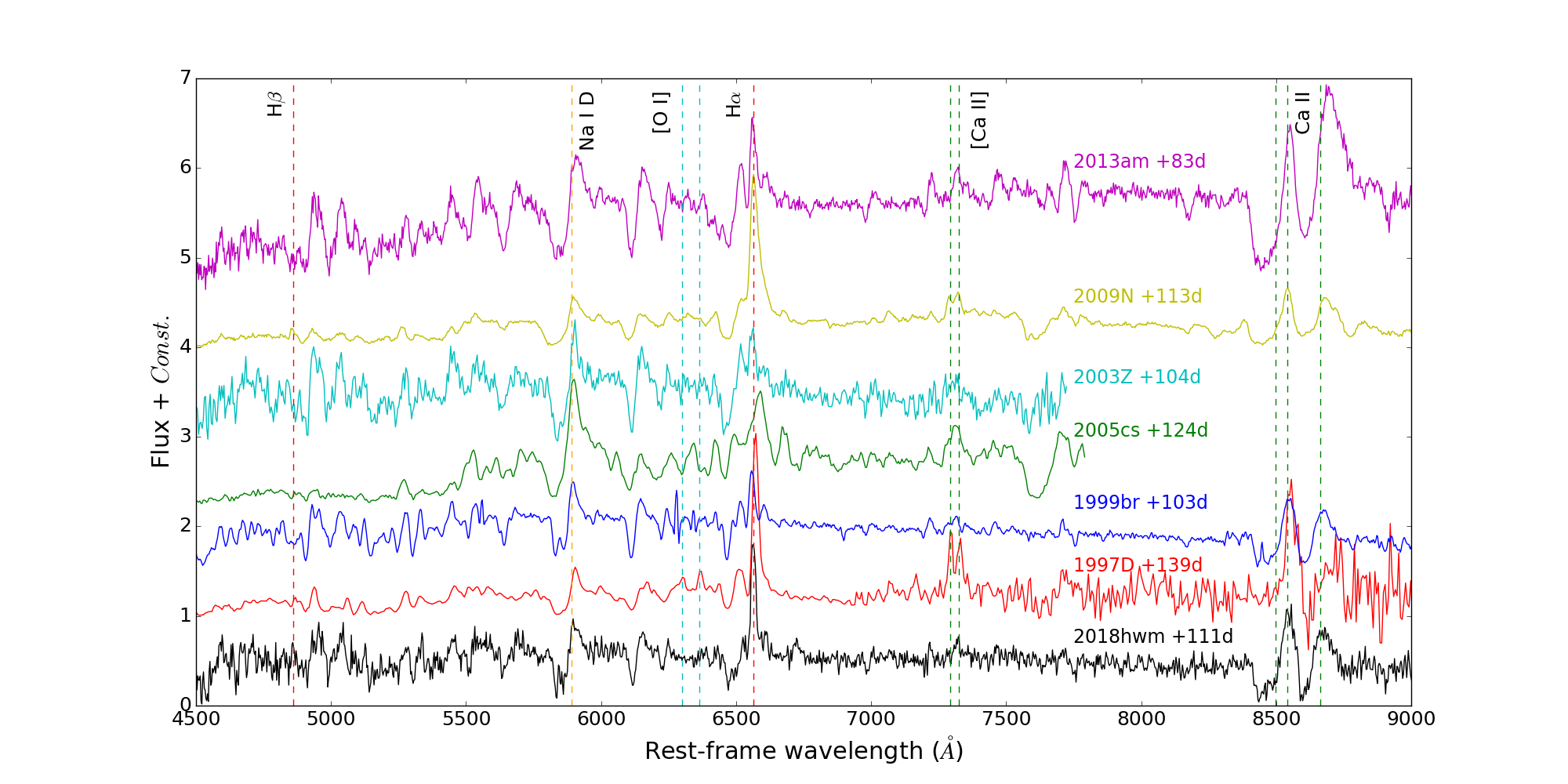}
\includegraphics[width=2\columnwidth]{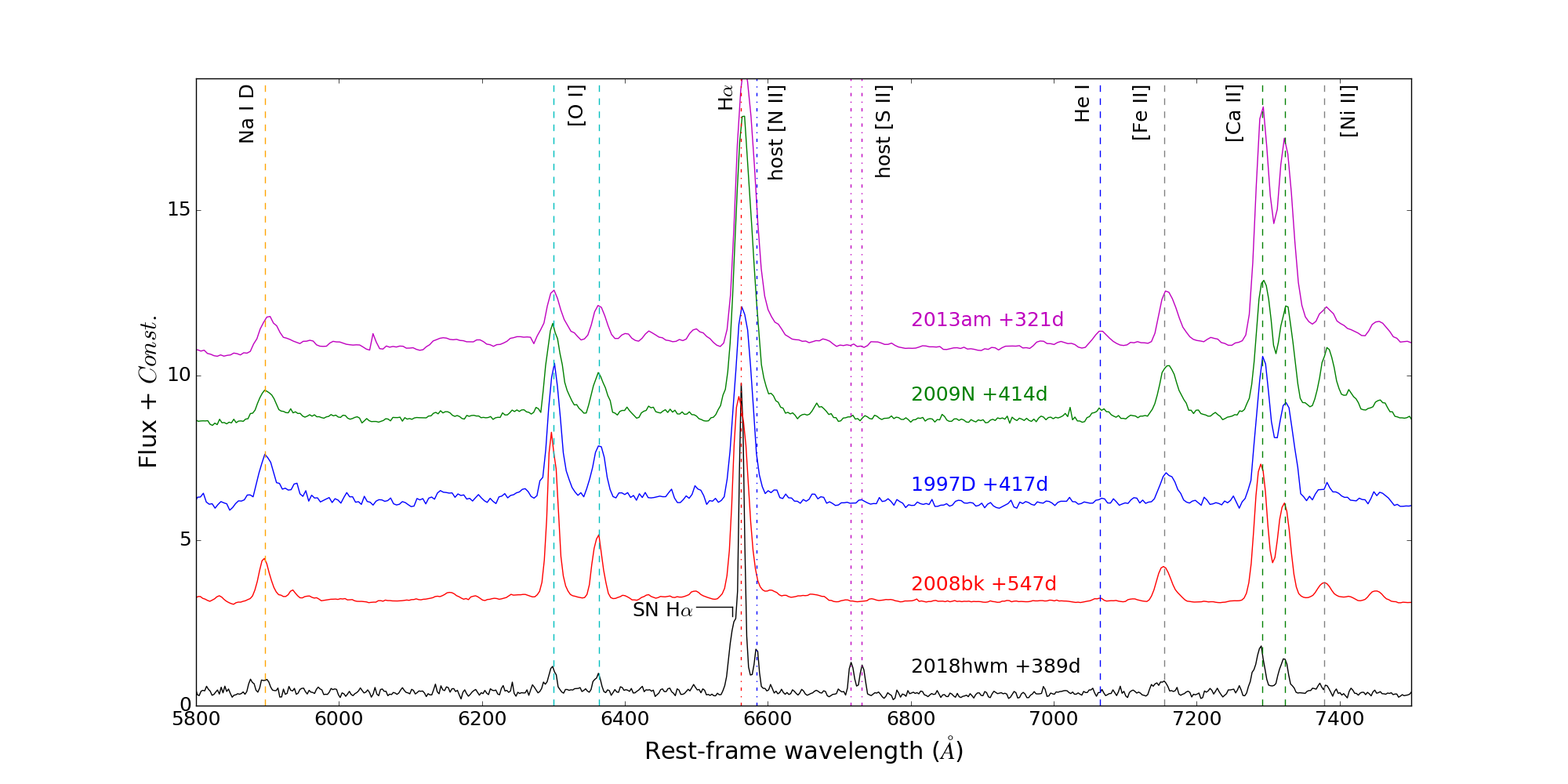}
\caption{Top: The last photospheric spectrum of SN 2018hwm compared with the faint IIP SNe 1997D, 1999br, 2005cs, 2003Z and the transitionals SNe 2009N and 2013am at similar phases.
Bottom: The nebular spectrum of SN 2018hwm is compared to those, taken at similar phases, of other faint IIP, SNe 1997D, 2008bk, 2009N and 2013am. Contaminant lines from the host are marked differently.}
\label{fig5}
\end{figure*}

In Figure \ref{fig5} (top) we compare the spectrum of SN 2018hwm taken at phase +113d, near to the end of the plateau, with those of both faint SNe IIP and transitional objects. The spectra of the comparison objects are taken close to the end of their plateaus. All these spectra are similar.

In Figure \ref{fig5} (bottom), we compare the nebular spectrum of SN 2018hwm with those of two under-luminous objects, SNe 1997D and 2008bk \citep[from][]{maguire}, and two transitional objects, SNe 2009N and 2013am. The H$\alpha$ profile in SN 2018hwm is remarkable, showing a blue-shifted, broader component from the SN beside the narrower, host contribution. The higher spectral resolution of GTC+OSIRIS allows to better resolve the [Ca II] doublet. Indeed, the emission lines of SN 2018hwm are among the narrowest observed in the spectra of faint Type IIP SNe. Some differences in the spectrum of SN 2018hwm are noticeable: the Na I D, [O I] doublet, [Fe II] and also [Ca II] lines are fainter and narrower. Transitional objects show a weak He I 7065 emission, which is not detected in the spectra of fainter SNe.

\section{Discussion} \label{discussion}
\subsection{Bolometric light curve} \label{bolometric}
\begin{figure}
\includegraphics[width=1.1\columnwidth]{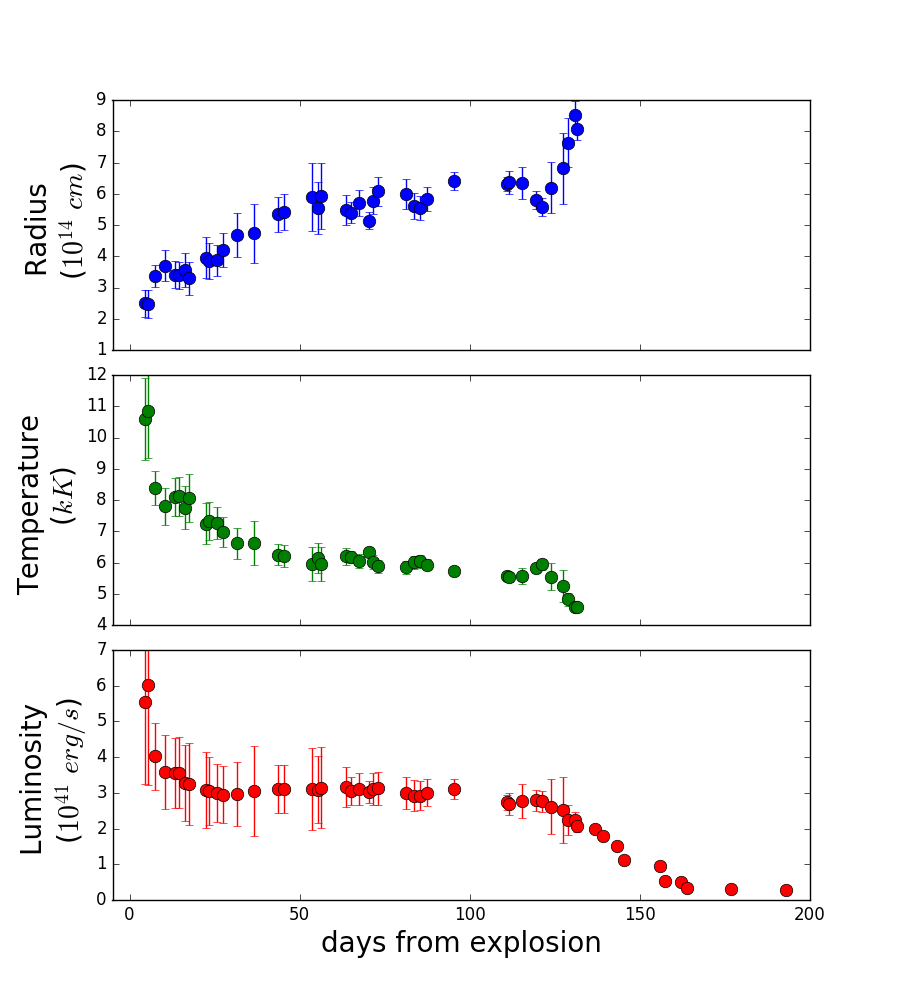}
\caption{Top: evolution of the radius of the photosphere, obtained by fitting the SED with a single black body. Center: evolution of the temperature of the photosphere. The radius and temperature evolutions are plotted only until +130 d, then the black body fit becomes unreliable. Bottom: bolometric light curve, calculated from the Stefan-Boltzmann law $L=4\pi r^2 \sigma T^4$. The phases are relative to the explosion epoch.} 
\label{fig6}
\end{figure}

We constructed the spectral energy distribution (SED) of SN 2018hwm from the available reddening-corrected photometry, spanning from $B$ to $K$ bands, and using the Sloan-$r$ as reference. When the photometric observation is missing in a given filter, its value is inferred from adjacent epoch photometry assuming a constant colour evolution. Then, we fitted the SED with a single black body curve, using the \texttt{curve\_fit}\footnote{https://docs.scipy.org/doc/scipy/reference/generated/ scipy.optimize.curve\_fit.html} tool of \textit{Python}. The free parameters of the fit are the radius, the temperature and the bolometric luminosity of the black body, that is derived by integrating the black body curve over all wavelenghts, and assuming the distance modulus value reported in Section 2.1. The errors of the free parameters are estimated from the covariance matrix provided by \texttt{curve\_fit}.
The bolometric light curve and the evolution of the radius and temperature of the black body are plotted in Figure \ref{fig6}.

The fit of the SED with a black body after the plateau end is not reliable, as the spectrum becomes dominated by emission lines rather than the continuum. Thus, we considered the evolution of the radius and temperature only until +130 d.
The black body radius starts to increase just after the explosion, levelling out to around 6$\times$10$^{14}$ cm (8600 AU) during the plateau. At the end of the plateau, the radius sharply increases, reaching nearly 9$\times$10$^{14}$ cm.
The temperature of the black body drops rapidly after the explosion, from 11000 to 8000 K in a few days. Later on, the temperature slowly decrease from 7500 to 5000 K.
The bolometric luminosity of SN 2018hwm shows an evolution similar to the one of the black body temperature, while also mirroring the $r$-band light curve. The mean bolometric luminosity during the plateau is 3$\times10^{41}$ erg s$^{-1}$, and drops by a factor of 10 when the object enters in the nebular phase.

\subsection{Hydrodynamical modelling} \label{hydro}
The physical properties of the progenitor of SN 2018hwm at the time of the explosion (the ejected mass $M_{ej}$, the progenitor radius at the explosion $R$ and the total explosion energy $E$) are derived using the same well-tested radiation-hydrodynamical modelling procedure that has been applied to other observed underluminous Type IIP SNe (e.g. \citealt{tomasella1}, \citealt{spiro2}, \citealt{takats2}, \citealt{takats3}, \citealt{pumo4}, \citealt{tomasella2}).
A complete description of this procedure is found in \cite{pumo4}. Here we recall its main features that can be summarized as follows.\par
\begin{itemize}
 \item[(i)] The SN progenitor's physical properties at the time of the explosion are constrained through the hydrodynamical modelling of all the main SN observables (i.e.~bolometric light curve, evolution of metal line velocities and the temperature at the photosphere), using a simultaneous $\chi^2$ fit of the observables against model calculations.
 \item[(ii)] The models are computed making use of the general-relativistic, radiation-hydrodynamics Lagrangian code presented in \cite{pumo2}. This code was specifically designed to simulate the evolution of the physical properties of SN ejecta and the behavior of the main SN observables during the entire post-explosive evolution (i.e.~from the breakout of the shock wave at the stellar surface up to the radioactive-decay phase), taking into account both the gravitational effects of the compact remnant and the heating effects due to the decays of the radioactive isotopes synthesized during the explosion. The four basic parameters guiding the post-explosion evolution of these models are $M_{ej}$, $R$, $E$ and the $^{56}$Ni mass, $M_{Ni}$, initially present in the ejecta of the models \citep[see also][]{pumo3}.
 \item[(iii)] The free model parameters of the $\chi^{2}$ fit are $M_{ej}$, $R$ and $E$. $M_{Ni}$ is instead held fixed and its value is set so as to reproduce the observed bolometric luminosity during the radioactive decay phase.
 \item[(iv)] The observational data taken at the earliest phases (i.e.~within the first $\sim$ 5 days after explosion for SN 2018hwm) are not included in the $\chi^{2}$ fit because the models could not accurately reproduce the early evolution of the main observables.
 \item[(v)] A preparatory analysis, aimed at determining the parameter space describing the SN progenitor at explosion, is performed with the semi-analytic code described in \cite{zampieri2}. This analysis guides the general-relativistic, radiation-hydrodynamics simulations that are more realistic but time consuming. Moreover the semi-analytic code is used to estimate the uncertainties on the free model parameters due to the $\chi^{2}$ fitting procedure. Possible systematic errors linked to the input physics (e.g.~the opacity treatment) and to the assumptions made in evaluating the observational quantities (e.g.~the adopted reddening or the adopted distance modulus) are not included.
\end{itemize}

\begin{figure*}
\includegraphics[width=1.3\columnwidth, angle=270]{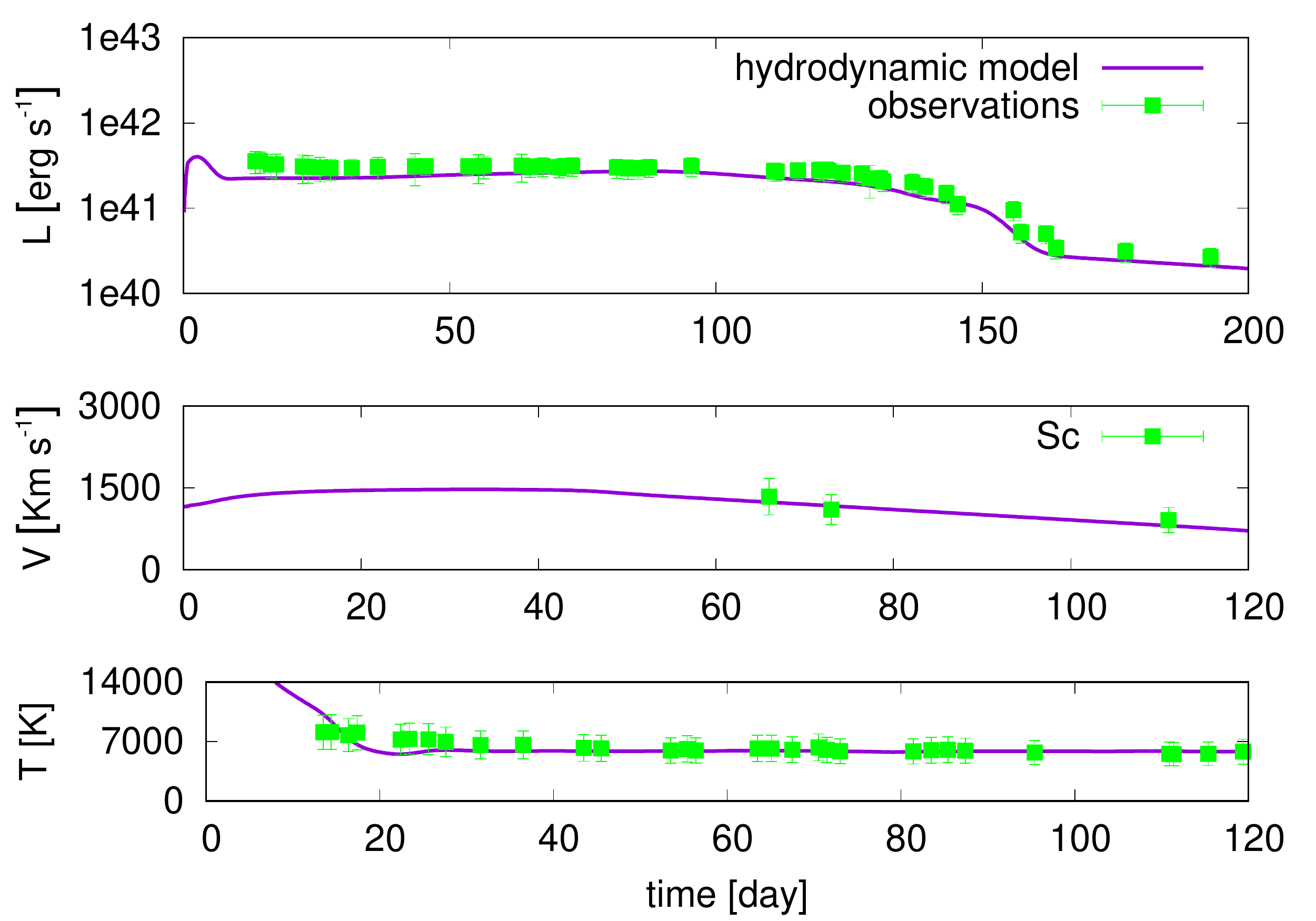}
\caption{Comparison of the evolution of the main observables of SN 2018hwm with the best-fit model computed with the general-relativistic, radiation-hydrodynamics code. The best-fit model parameters are: total energy $0.075$ foe, radius at explosion $5.9\times10^{13}$ cm, and ejected mass 8.0~$M_{\odot}$. Top, middle, and bottom panels show the bolometric light curve, the photospheric velocity, and the black-body photospheric temperature as a function of time, respectively. To better estimate the photosphere velocity from observations, we use the minimum of the profile of the Sc lines, as the Sc lines are generated just above the photosphere.} 
\label{fig:bestfitmodel}
\end{figure*}

Based on the adopted explosion epoch (cfr. Sect. \ref{lightcurves}) and bolometric luminosity (cfr. Sect. \ref{bolometric}), we find the best-fitting model shown in Figure \ref{fig:bestfitmodel}. In particular, the inferred best-fit model has a kinetic plu thermal energy of 0.075 foe, radius at explosion of $5.9\times10^{13}$ cm ($\sim$845~$R_{\odot}$), $M_{Ni}\simeq0.0085$ and ejected mass of 8.0~$M_{\odot}$. Adding the mass of the compact remnant ($\sim$1.3-2~$M_{\odot}$) to that of the ejected material, we obtain a total stellar mass at explosion of $\sim$9.3-10.0~$M_{\odot}$. To estimate the initial mass of the progenitor on the MS, we assume a typical (i.e. not enhanced by rotation) mass loss during the pre-SN evolution $\lesssim$ 0.1-0.9 $M_{\odot}$ (see \citealt{pumo4} and references therein for details). Hence, the initial progenitor mass is in the range 9.4-10.9~$M_{\odot}$. The estimated errors on the free model parameters due to the $\chi^{2}$ fitting procedure are about 15\% for $M_{ej}$ and $R$, and 30\% for $E$.

\cite{popov} and \cite{kasen} derived scaling relations for the properties of SNe IIP, and concluded that the plateau duration is correlated to the envelope mass. While in SN 2018hwm the plateau is very long (nearly 140 days), the ejecta mass is not higher than the expectation from the analytical modelling of the explosion. The discrepancy is explained by the extremely low explosion energy (more than one order of magnitude smaller than the typical value derived by \cite{kasen} ($\sim0.9\times10^{51}$ ergs). The low kinetic energy of the ejecta, and the consequent low expansion velocities, generate a slow-moning recombination. As a consequence, the plateau has a longer duration than that usually observed in SNe IIP. \cite{pumo4} noted that the main regulator of the distribution of IIP SNe properties (such as luminosity and $^{56}$Ni mass) is in fact the ratio between the explosion energy and ejecta mass ($E/M_{ej}$).
For SN 2018hwm, we evaluate the value of this ratio to $E/M_{ej}$=0.0094 foe/$M_{\odot}$, which is among the lowest calculated for a Type IIP SN (\citealt{pumo4}, \citealt{lisakov2} and references therein), and is nearly half of what is derived for other low luminosity SNe IIP with known RSG progenitors.

\subsection{Explosion and progenitor scenario}
The derived explosion energy of SN 2018hwm (0.075 foe), when compared to the theoretical models \citep{sukhbold,burrows,limongi}, is extremely low, even accounting for the error bar. Standard SN explosion simulations predict explosion energies above 0.2$\times10^{51}$ erg, and up to 2$\times10^{51}$ erg. Although it is difficult to explain such a low energy with a standard core-collapse SN, it is still possible. For example, SN 1054 (a.k.a. `the Crab event') is believed to be the outcome of a ECSN \citep{kitaura} with an explosion energy of 0.1 foe \citep{sukhbold}. In addition, using the scaling relations from \cite{popov}, \cite{muller} are able to obtain $E$ below 0.1 foe for a few normal Type II SNe.

The $^{56}$Ni synthesized in SNe, which is the only source powering the late-time luminosity after the plateau stage, is largely formed by the explosive burning of oxygen and silicon.
To explain the ejection of a small amount of $^{56}$Ni in LL IIP, three main explosion scenarios have been formulated:
(i) The fall-back of most of the material from the outer, Ni-rich region of a high mass (25-40 $M_{\odot}$) progenitor star onto its newly formed degenerate core (\citealt{woosley2}, \citealt{zampieri1}; \citealt{moriya}). This phenomenon would naturally produce a low luminosity explosion event, and the apparent formation of small quantities of $^{56}$Ni, as it falls back on to the newly born black hole.
(ii) The detonation of a low mass (8–10 $M_{\odot}$) progenitor as a Fe core-collapse SN, in which very little $^{56}$Ni is synthesized (\citealt{nomoto1}, \citealt{woosley1}, \citealt{chugai}).
(iii) A third explosion scenario, that can explain the observed properties of faint IIP SNe (i.e. massive ejecta, low $^{56}$Ni mass produced and low energetic events) is the explosion of a moderate-mass star, with O-Ne-Mg core, via Electron-Capture SN (ECSN; \citealt{nomoto2}, \citealt{ritossa}, \citealt{heger}, \citealt{kitaura}; \citealt{takahashi}).
The theory of stellar evolution predicts stars with a MS mass of 8 to 11 $M_{\odot}$ to become SAGB stars which may lead to ECSNe \citep[e.g.][]{tominaga}.
\cite{kitaura} and \cite{eldridge} noted that a SAGB progenitor will have little oxygen and silicon surrounding the core, and hence we may expect a SAGB star to produce a low mass of ejected $^{56}$Ni.

\cite{franson} suggested that the ratio $\Re$ between the luminosities of the [Ca II] $\lambda\lambda$7291,7324 and [O I] $\lambda\lambda$6300,6364 can be a good diagnostic for the main-sequence mass of the precursor star. We calculated this ratio for SN 2018hwm in the nebular spectrum (phase +389 d), and obtained $\Re\sim$2.7. A value of $\Re$ close to 3 is consistent with a low main-sequence progenitor mass, as estimated for SN 2005cs \citep[7–13 $M_{\odot}$; \citealt{maund1}, \citealt{li1}, \citealt{takats1} and][]{eldridge}.

The inferred values of the best-fit model parameters are consistent with an ECSN involving a SAGB star with initial mass close to the upper limit of the mass range typical of this class of stars \citep[for further details see e.g.][]{pumo1}.

The best-fit model parameters may be also consistent with an iron core-collapse SN with a very low-energy explosion involving a low-mass red/yellow supergiant star. However, the inferred value of $E$ seems to be too low to consider this scenario as the most reliable, although it cannot be ruled out.
Lacking information on the temperature and colour of the progenitor from pre-explosion images (as was obtained for SNe 2005cs and 2008bk), we cannot distinguish an SAGB from an RSG progenitor.
The absence of signs of interaction with a pre-existing CSM, particularly in the spectra, has two possible explanations: (i) the progenitor experienced only very minor mass loss before the core-collapse, which is not expected from a star in the SAGB phase, or (ii) the progenitor was a SAGB that had recently ($\sim 10^4$ years, see Table 1 from \citealt{pumo1}) entered the Thermal Pulses phase. In the latter case, the star would have retained most of the envelope. This is ejected only in the final explosive event, such that a dense CSM is not present.

The progenitor mass of SN 2018hwm obtained from hydrodynamical simulations is in agreement with the value found for many subluminous SNe. For example, \cite{spiro2} propose a progenitor masses range of 10–15 $M_{\odot}$ for their sample of LL SNe IIP.
\cite{utrobin1} modelled the light curves and spectra of SN 2003Z, obtaining a high ejecta mass of 14 $M_{\odot}$, that leads to an estimated initial mass of 16 $M_{\odot}$ but a small progenitor radius of 230 $R_{\odot}$. \cite{pumo4} found a similar value of 260 $R_{\odot}$. The small radius may favour a yellow supergiant as progenitor, instead of a red one, and it has been hypothesized for the transitional event SN 2009N \citep{takats2}.

As mentioned before, we cannot provide conclusive arguments supporting one of the two explosion scenarios. As some observational properties seem to favour the ECSN and others the Fe-CC SN, we can here summarise their strong and weak arguments. The arguments in favour of an ECSN scenario for SN 2018hwm are:
\begin{itemize}
\item the low explosion energy ($7.5\times10^{49}$ erg),
\item the low synthesized $^{56}$Ni mass (0.0085 $M_{\odot}$),
\item the low initial progenitor mass (between 9 and 11 $M_{\odot}$),
\item the absence of the He I $\lambda$7065 line \citep{jerkstrand3}.
\end{itemize}

On the other hand, the more canonical Fe-CCSN scenario can also explain some of the observed properties, such as:
\begin{itemize}
\item a star with a radius of 845 $R_{\odot}$ is also compatible with a RSG,
\item the large amount of $^{40}$Ca synthesised (see below),
\item other LL IIP with similar parameters, like SN 2005cs \citep{eldridge} and SN 2008bk (\citealt{mattila1}, \citealt{vandyk2}), were confirmed to have RSG progenitors.
\end{itemize}
Two possible arguments against the Fe-CCSN scenario are:
\begin{itemize}
\item SN 2018hwm should be an extremely faint Fe-CCSN in order to be able to explain the energy and M($^{56}$Ni) values,
\item the relatively low progenitor mass at ZAMS.
\end{itemize}

\cite{lisakov2} investigated whether a low-energy explosion of high-mass (12 to 27 $M_{\odot}$) RSGs could reproduce the observed properties of LL SNe. Their results were in contrast with the observations, as they predict bluer colours and faster declining light curves, rather than those observed during the plateau. Also, the complete fall-back of the CO core prevents the ejection of any $^{56}$Ni, whereas LL SNe IIP produce a small amount of it. The results support a scenario involving low- to intermediate-mass progenitors, pointing to low-energy explosions of RSG or SAGB stars.

\subsection{Nucleosynthesis modeling}
\begin{figure*}
\includegraphics[width=.7\columnwidth]{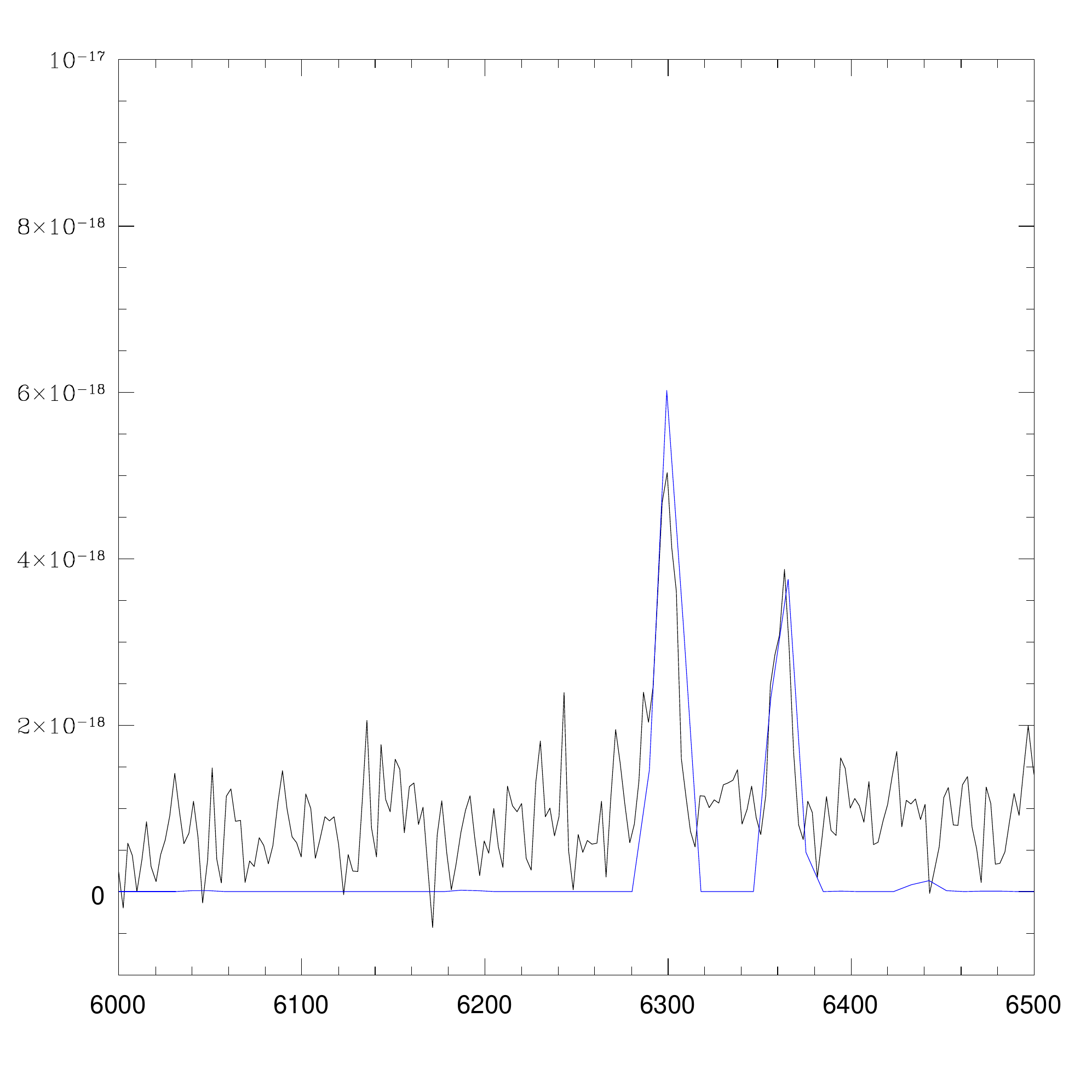}\includegraphics[width=.7\columnwidth]{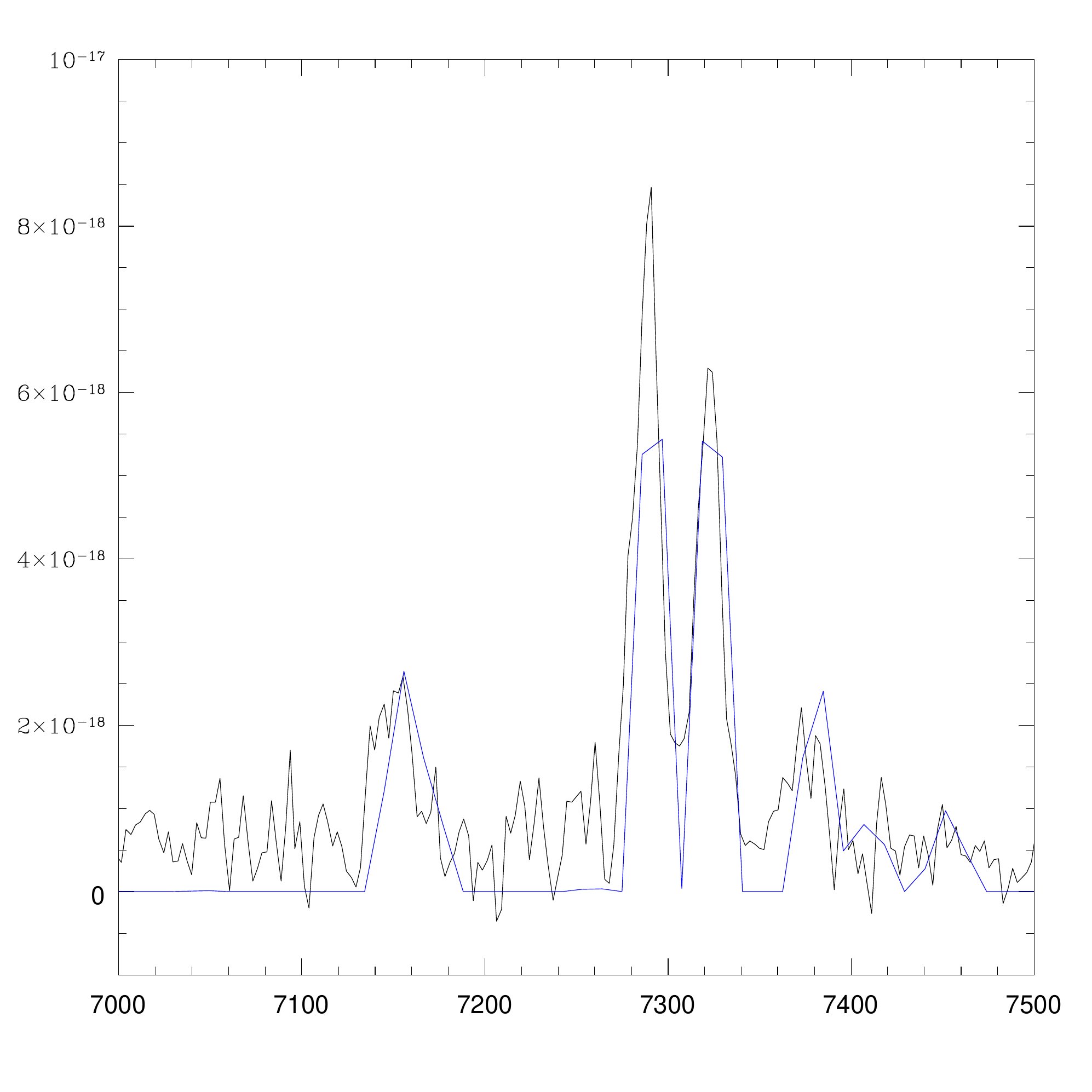}\includegraphics[width=.7\columnwidth]{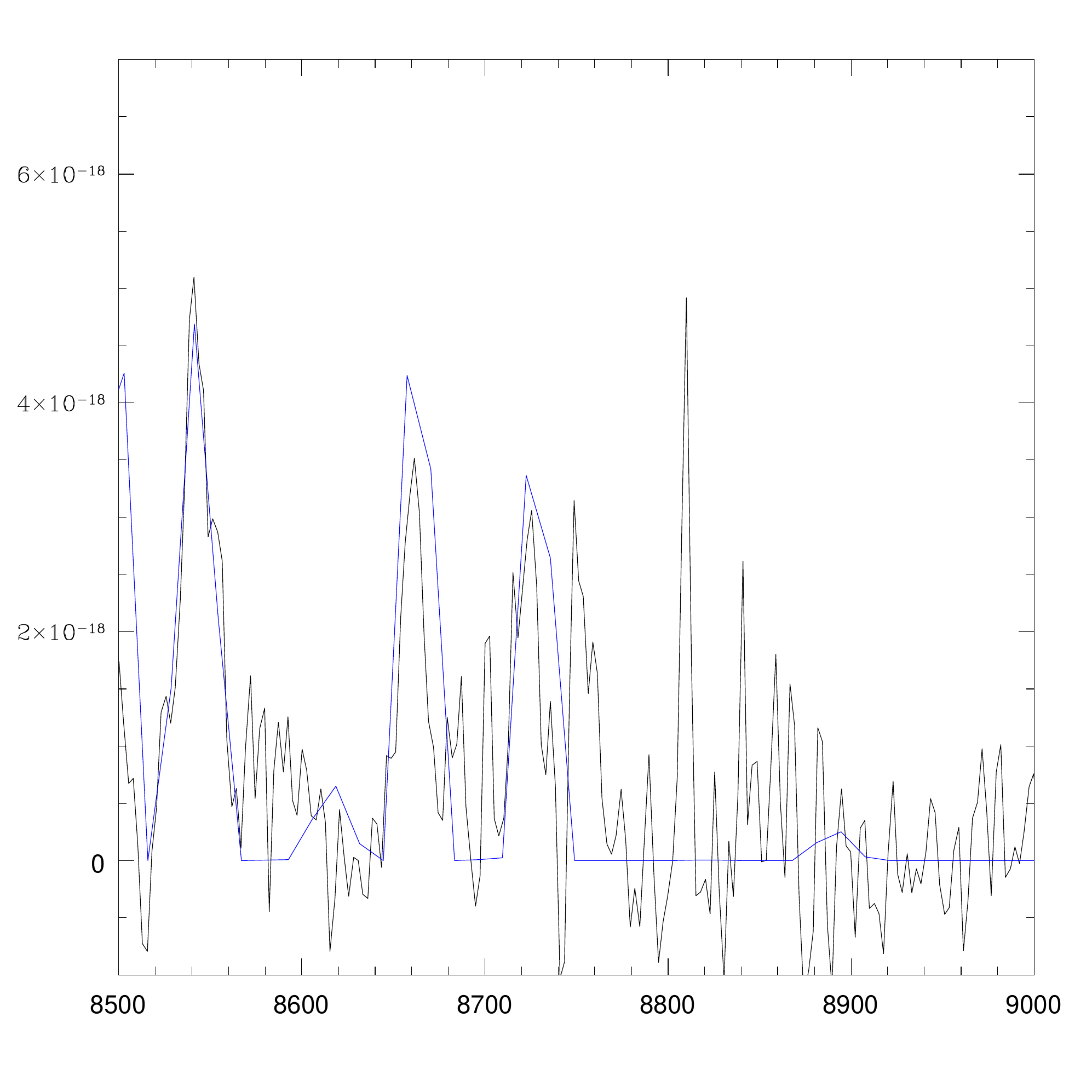}
\caption{Three snapshots of the modeling (in blue) of the nebular spectrum of SN 2018hwm (in black), fitting the three most important features of the spectrum. From left to right: [O I] doublet, [Ca II] doublet and NIR Ca II triplet. From the model we confirmed the ejected M($^{56}$Ni) of 0.002 $M_{\odot}$ derived from the hydrodynamical simulation, and estimated the ejected masses of $^{16}$O, $^{12}$C and $^{40}$Ca.} 
\label{fig8}
\end{figure*}

In order to establish the properties of the core of the exploding star, we modelled the nebular spectrum with a non-local thermodynamic equilibrium (NLTE) code. The code is based on the assumptions set forth by \cite{axelrod}, and it has been described and used in several papers (e.g. \citealt{mazzali1,mazzali2}). The ejecta are heated and ionised by impact with the products of the radioactive decay chain $^{56}$Ni-$^{56}$Co-$^{56}$Fe. Cooling occurs mostly via forbidden line emission. If we exclude the H and He envelopes, the heavier elements that are found in the CO core can be modelled independently. Although Fe lines are not visible in the noisy spectrum, we can set the $^{56}$Ni mass to the value obtained from the light curve, adapt the M($^{56}$Ni) by verifying that the resulting Fe emission in the region of 4000-5500 \AA~does not exceed the observed flux, and test the masses of some of the emitting elements. It is immediately clear that all emission lines from heavy elements are very narrow. A typical nebular velocity of 500 kms$^{-1}$ appropriately describes their width, so we assume that is the size of the emitting core. The strongest lines we address are [O I] 6300, 6363, [Ca II] 7321, Ca II IR, Mg I] 4571, [C II] 8600. 

Three snapshots of the comparison between the model spectrum with the observed one, around the [O I], [Ca II] doublets and NIR Ca II triplet, are shown in Figure \ref{fig8}.
By approximating to the blue emission we obtained a M($^{56}$Ni) of 0.002 $M_{\odot}$. The inferred oxygen and carbon masses are low, M($^{16}$O)$\approx 0.02$ $M_{\odot}$ and M($^{12}$C)$\approx 0.005 M_{\odot}$, respectively.
The newly synthesized M($^{16}$O) found for SN 2018hwm is very close to that obtained for SN 2005cs (0.016 $M_{\odot}$, \citealt{jerkstrand3}).
This agrees with the findings that MS stars in the 8-12 $M_{\odot}$ range synthesize low amount of oxygen (a few $10^{-2}$ $M_{\odot}$, \citealt{nomoto2}; \citealt{woosley1}; \citealt{woosley3}; \citealt{chugai}).

We require a high ejected calcium mass, $\sim$0.3 $M_{\odot}$, to match the strong observed emission. The Ca lines are not broader than the O line, so it is unlikely that any other region contributes significantly to the emission. In order to achieve the correct ratio of the Ca II lines we need to use a rather strong degree of clumping (filling factor 0.02-0.05), which likely reflects the actual conditions in the young SN remnant.
A high M($^{40}$Ca) is problematic within the ECSN scenario, as theoretical nucleosynthesis calculations predict much lower values ($<$0.01 $M_{\odot}$, \citealt{nomoto3}). 
As a reference, the 12 $M_{\odot}$ model of \cite{maguire} for SN 2008bk synthesised 2.4E-03 $M_{\odot}$ of calcium. However, a high Ca abundance has been observed in a number of so-called `Ca-rich' transients \citep[e.g.][with M(Ca)=0.135 $M_{\odot}$]{perets1}. Those are more likely to be He-shell detonations on white dwarfs, although \cite{kawabata} suggest two scenarios to explain the Ca-rich Type Ib SN 2005cz: the core-collapse of a low-mass 8-12 $M_{\odot}$ star in a binary system, or an ECSN explosion induced by the merging of a ONeMg and a He white dwarf.

We can conclude that the ejected mass is quite small, and the explosion energy is very low, but it is difficult to distinguish between ECSN and a low-mass Fe-CCSN scenario based solely on the nucleosynthesis. For example, \cite{wanajo} proved that the collapse of a ONeMg core and the least massive Fe-core lead to similar nucleosynthesis.
Hence, with the available data, we cannot securely discriminate between EC and Fe-CC scenarios.

The number of LL SNe IIP with extensive coverage both in photometry and spectroscopy allowing detailed modelling is still limited. The number of LL SNe IIP with progenitor information inferred from the direct analysis of pre-explosion \textit{HST} images is even lower. However, with the available sample, there is a growing evidence that the masses of the progenitors of LL SNe IIP are too low to comfortably match the expectations for a fall-back SN scenario, whereas relatively low-mass RSG leading to an Fe core-collapse or SAGBs producing EC SN explosions are viable alternatives. 

\section*{Acknowledgments}
\begin{small}
MLP acknowledges support from the plan "programma ricerca di ateneo UNICT 2020-22 linea 2" of the Catania University. HK was funded by the Academy of Finland projects 324504 and 328898. GP acknowledge support by the Ministry of Economy, Development, and Tourism’s Millennium Science Initiative through grant IC120009, awarded to The Millennium Institute of Astrophysics, MAS. OR acknowledges support from CONICYT PAI/INDUSTRIA 79090016.
IRAF is distributed by the National Optical Astronomy Observatories, which are operated by the Association of Universities for Research in Astronomy, Inc., under cooperative agreement with the National Science Foundation.
Based on observations obtained with the Samuel Oschin 48-inch Telescope at the Palomar Observatory as part of the Zwicky Transient Facility project. ZTF is supported by the National Science Foundation under Grant No. AST-1440341 and a collaboration including Caltech, IPAC, the Weizmann Institute for Science, the Oskar Klein Center at Stockholm University, the University of Maryland, the University of Washington, Deutsches Elektronen-Synchrotron and Humboldt University, Los Alamos National Laboratories, the TANGO Consortium of Taiwan, the University of Wisconsin at Milwaukee, and Lawrence Berkeley National Laboratories. Operations are conducted by COO, IPAC, and UW.\\
Based on observations made with the Nordic Optical Telescope, operated by the Nordic Optical Telescope Scientific Association, and with the Liverpool Telescope, operated on the island of La Palma by Liverpool John Moores University, with financial support from the UK Science and Technology Facilities Council. Both telescopes are located at the Spanish Observatorio del Roque de los Muchachos, La Palma, Spain, of the Instituto de Astrofisica de Canarias.
The ALFOSC instrument is provided by the Instituto de Astrofisica de Andalucia (IAA) under a joint agreement with the University of Copenhagen and NOTSA.
The NUTS program is funded in part by the IDA (Instrument Centre for Danish Astronomy).
Based on observations made with the Gran Telescopio Canarias (GTC), installed at the Spanish Observatorio del Roque de los Muchachos of the Instituto de Astrofísica de Canarias, in the island of La Palma.
Based on observations collected at the European Organisation for Astronomical Research in the Southern Hemisphere, Chile, as part of ePESSTO (the extended Public ESO Spectroscopic Survey for Transient Objects Survey) ESO program ID 199.D-0143(M).
Based on observations at Cerro Tololo Inter-American Observatory, which is managed by the Association of Universities for Research in Astronomy (AURA) under a cooperative agreement with the National Science Foundation.
The Dark Energy Camera (DECam) was constructed by the Dark Energy Survey (DES) collaboration.
We acknowledge the usage of the HyperLeda database (http://leda.univ-lyon1.fr).

\section*{Data availability}
The data underlying this article are available in the article and in its online supplementary material.

\end{small}

\appendix
\setcounter{table}{0}
\renewcommand{\thetable}{A\arabic{table}}
\section*{Appendix: Photometry tables}
\begin{table*}
\caption{Sloan $griz$ AB magnitudes of SN 2018hwm.}
\label{tab4}
\begin{tabular}{ccccccl}
\hline\hline
Date & MJD & $g$ & $r$ & $i$ & $z$ & Instrument \\
\hline
2018/11/01 & 58423.49 & $>$20.9 & $>$20.9 & - & - & ZTF\\
2018/11/04 & 58426.48 & 18.32 0.05 & 18.62 0.03 & - & - & ZTF\\
2018/11/07 & 58429.49 & 18.38 0.06 & 18.48 0.05 & - & - & ZTF\\
2018/11/08 & 58430.41 & 18.32 0.05 & - & - & - & ZTF\\
2018/11/09 & 58431.25 & 18.36 0.01 & 18.40 0.01 & - & - & DECAM \\
2018/11/10 & 58432.48 & 18.60 0.04 & 18.50 0.04 & - & - & ZTF\\
2018/11/13 & 58435.52 & - &  18.53 0.09 & - & - & ZTF\\
2018/11/16 & 58438.49 & 18.75 0.05 & 18.60 0.04 & - & - & ZTF\\
2018/11/17 & 58439.41 & 18.74 0.05 & 18.59 0.03 & - & - & ZTF\\
2018/11/19 & 58441.44 & 18.86 0.04 & 18.65 0.03 & - & - & ZTF\\ 
2018/11/20 & 58442.38 & - & 18.76 0.08 & - & - & ZTF\\
2018/11/25 & 58447.42 & - & 18.60 0.08 & - & - & ZTF\\
2018/11/26 & 58448.43 & - & 18.65 0.08 & - & - & ZTF\\
2018/12/09 & 58461.47 & 19.37 0.15 & 18.56 0.03 & - & - & ZTF\\
2018/12/13 & 58465.51 & 19.35 0.08 & 18.71 0.04 & - & - & ZTF\\
2018/12/16 & 58468.46 & 19.34 0.06 & 18.58 0.03 & - & - & ZTF\\
2018/12/20 & 58472.45 & 19.59 0.14 & 18.77 0.03 & - & - & ZTF\\ 
2018/12/28 & 58480.45 & 19.45 0.17 & 18.53 0.12 & - & - & ZTF\\
2018/12/29 & 58481.43 & - & 18.48 0.04 & - & - & ZTF\\
2019/01/04 & 58487.44 & 19.27 0.10 & 18.64 0.04 & - & - & ZTF\\
2019/01/07 & 58490.09 & - & 18.57 0.11 & - & - & ALFOSC\\
2019/01/12 & 58495.26 & 19.44 0.09 & - & - & - & ZTF\\ 
2019/01/15 & 58498.00 & - & 18.44 0.04 & - & - & ALFOSC\\
2019/01/26 & 58509.26 & 19.33 0.13 & 18.64 0.06 & - & - & ZTF\\
2019/02/09 & 58523.23 & 19.44 0.12 & 18.65 0.03 & - & - & ZTF\\
2019/02/12 & 58526.31 & 19.34 0.23 & 18.69 0.06 & - & - & ZTF\\   
2019/02/20 & 58534.28 & - & 18.57 0.16 & - & - & ZTF\\ 
2019/02/21 & 58535.95 & - & 18.64 0.06 & - & - & ALFOSC\\
2019/02/24 & 58538.25 & 19.81 0.24 & - & - & - & ZTF\\ 
2019/02/26 & 58540.93 & 19.65 0.04 & 18.68 0.02 & 18.53 0.02 & 18.52 0.04 & IO:O\\
2019/03/02 & 58544.88 & 19.48 0.05 & 18.70 0.02 & 18.56 0.03 & 18.54 0.07 & IO:O\\
2019/03/06 & 58548.03 & - & - & 18.70 0.08 & 18.33 0.11 & APOGEE\\
2019/03/06 & 58548.84 & 19.63 0.06 & 18.71 0.03 & 18.54 0.03 & 18.56 0.05 & IO:O\\
2019/03/11 & 58553.89 & 19.82 0.06 & 18.84 0.02 & 18.65 0.03 & 18.57 0.04 & IO:O\\
2019/03/13 & 58555.85 & 20.30 0.04 & 18.91 0.03 & 18.80 0.02 & 18.62 0.02 & IO:O\\
2019/03/15 & 58557.02 & - & 18.90 0.10 & - & - & APOGEE\\
2019/03/16 & 58558.23 & - &  18.98 0.11 & - & - & ZTF\\ 
2019/03/17 & 58559.03 & - & - & 19.04 0.16 & 18.64 0.30 & APOGEE\\
2019/03/19 & 58561.87 & 20.16 0.29 & 19.35 0.18 & 18.97 0.09 & 18.79 0.12 & IO:O\\
2019/03/25 & 58567.17 & 20.95 0.20 & - & - & - & ZTF\\ 
2019/04/02 & 58575.17 & 20.98 0.23 & - & - & - & ZTF\\ 
2019/04/07 & 58580.89 & - & - & 20.52 0.04 & 20.33 0.05 & ALFOSC\\
2019/04/13 & 58586.97 & 21.37 0.36 & 20.51 0.06 & 20.82 0.12 & 20.69 0.14 & IO:O\\
2019/04/15 & 58588.89 & 21.49 0.06 & 20.72 0.04 & 20.87 0.04 & 20.79 0.06 & ALFOSC\\
2019/04/25 & 58598.17 & 21.25 0.26 & 21.12 0.19 & - & - & ZTF\\ 
2019/04/28 & 58601.87 & 22.69 0.15 & 21.25 0.04 & 20.75 0.05 & 20.94 0.07 & ALFOSC\\
2019/05/14 & 58617.87 & - & 21.28 0.05 & 20.94 0.07 & - & ALFOSC\\
2019/10/04 & 58760.25 & - & 22.24 0.09 & - & - & ALFOSC\\
2019/11/27 & 58814.20 & - & 22.85 0.13 & - & - & OSIRIS\\
\hline
\end{tabular}
\end{table*}

\begin{table}
\caption{Johnson $BVRI$ Vega magnitudes of SN 2018hwm.}
\label{tab5}
\begin{tabular}{ccccccl}
\hline\hline
Date & MJD & $B$ & $V$ & $R$ & $I$ & Instrument \\
\hline
2018/11/08 & 58430.33 & - & 18.43 0.03 & - & - & EFOSC\\
2019/01/10 & 58493.23 &	19.90 0.06 & 18.70 0.03 & - & - & ANDICAM\\
2019/01/15 & 58498.24 & 19.96 0.07 & 18.72 0.04 & - & 17.93 0.03 & ANDICAM\\
2019/01/24 & 58507.19 & 20.04 0.10 & 18.83 0.07 & - & 17.90 0.04 & ANDICAM\\
2019/02/01 & 58515.19 & 20.20 0.07 & 18.76 0.03 & - & 17.96 0.04 & ANDICAM\\
2019/02/03 & 58517.27 & 20.22 0.08 & 18.79 0.07 & 18.23 0.06 & 17.89 0.08 & Fairchild\\
2019/02/08 & 58522.15 & 20.17 0.06 & 18.81 0.05 & - & 17.92 0.08 & ANDICAM\\
2019/02/16 & 58530.14 & 20.26 0.10 & 19.06 0.16 & - & 17.92 0.09 & ANDICAM\\
2019/02/22 & 58536.23 & - & 19.03 0.06 &  18.44 0.05 & - & Fairchild\\
2019/02/23 & 58537.15 & - & 18.93 0.05 &  18.38 0.04 & 18.03 0.03 & ANDICAM\\
2019/02/26 & 58540.89 & 20.38 0.08 & 19.13 0.03 & - & - & IO:O\\
2019/03/02 & 58544.88 & 20.39 0.09 & 19.09 0.04 & - & - & IO:O\\
2019/03/06 & 58548.12 & - & 19.05 0.04 &  18.49 0.04 & 18.11 0.10 & ANDICAM\\
2019/03/06 & 58548.85 & 20.72 0.11 & 19.53 0.06 & - & - & IO:O\\
2019/03/11 & 58553.90 & 20.77 0.13 & 19.47 0.05 & - & - & IO:O\\
2019/03/15 & 58557.92 & 20.96 0.21 & 19.54 0.09 & - & - & IO:O\\
2019/03/19 & 58561.87 & - & 19.72 0.23 & - & - & IO:O\\
2019/03/22 & 58564.11 & - & 19.82 0.15 & 19.19 0.09 & 18.90 0.15 & Fairchild\\
2019/04/03 & 58576.06 & - & 20.60 0.14 & 20.10 0.10 & - & ANDICAM\\
2019/04/06 & 58579.07 & - & 20.65 0.11 & 20.08 0.14 & - & ANDICAM\\
2019/04/09 & 58582.05 & - & 20.70 0.14 & 20.09 0.13 & - & ANDICAM\\
2019/04/15 & 58588.88 & - & 21.40 0.11 & 20.46 0.05 & - & ALFOSC\\
2019/04/28 & 58601.86 & - & 21.37 0.08 & - & - & ALFOSC\\
2019/05/14 & 58617.87 & - & 21.36 0.11 & - & - & ALFOSC\\
\hline
\end{tabular}
\end{table}

\begin{table}
\caption{NIR $JHKs$ Vega magnitudes of SN 2018hwm.}
\label{tab6}
\begin{tabular}{cccccl}
\hline\hline
Date & MJD & $J$ & $H$ & $Ks$ & Instrument \\
\hline
2019/01/11 & 58494.21 & 17.36 0.03 & 17.08 0.02 & 17.01 0.05 & NOTCAM \\
2019/01/29 & 58512.99 & 17.43 0.02 & 17.13 0.02 & 16.91 0.03 & NOTCAM \\
2019/03/04 & 58546.93 & 17.80 0.03 & 17.40 0.03 & 17.21 0.05 & NOTCAM \\
2019/03/21 & 58563.97 & 19.09 0.08 & 18.39 0.07 & - 		 & NOTCAM \\
\hline
\end{tabular}
\end{table}

\end{document}